\documentstyle[prl,aps,epsf,cite]{revtex}
\input epsf.sty
\begin{document}
\twocolumn[\hsize\textwidth\columnwidth\hsize\csname@twocolumnfalse%
\endcsname
\title{Superconductivity in electron-doped cuprates:\\
Gap shape change and symmetry crossover with doping}
\author{Francisco Guinea$^{\dag}$,
Robert S. Markiewicz$^{\ddag}$ and Mar\'{\i}a A. H. Vozmediano$^{\star}$.}
\address{
$^{\dag}${I}nstituto de Ciencia de Materiales de Madrid,
CSIC, Cantoblanco, E-28049 Madrid, Spain \\
$^{\ddag}$Physics Department, Northeastern University, Boston MA 02115,
USA\\
$^{\star}$ Departamento de Matem\'aticas,
Universidad Carlos III de Madrid, 28911 Legan\'es,
Madrid, Spain.}
\date{\today}
\maketitle
\begin{abstract}The Kohn-Luttinger mechanism for superconductivity
is investigated in a model for the  electron doped cuprates.
The symmetry of the order  parameter of the superconducting phase
is determined as a function of the geometry of the Fermi surface together
with the structure of the electron-hole susceptibility. It is found to
remain d$_{x^2-y^2}$ wave within a large doping range.  The {\it shape} of
the gap anisotropy evolves with doping, with the maximum gap moving away
from $(\pi ,0)$, in good agreement with recent experiments.  As the shift
of the maximum increases, a crossover to $d_{xy}$-symmetry is found.
\end{abstract}

]
\narrowtext

\section{Introduction}

The single band Hubbard model with longer range hoppings -- (t-t')
Hubbard model -- is being widely accepted as a minimal model to describe
the physics of the high-$T_c$ cuprates \cite{and}.  It gives most of the
qualitative aspects of the phase diagram: antiferromagnetism near half
filling, superconductivity, pseudogap, and striped phases.
The greatest effort has previously been centered
in understanding hole-doped compounds
as they possess the highest available $T_c$. The complication
of the phase diagram in the underdoped to optimally doped
regimes, partially due to the inhomogeneous structures and
the proximity of the Van Hove singularity,  has recently renewed
interest in electron doped compounds \cite{nparm}.

Although the estimated values of the Hubbard interaction $U$
lie in the intermediate regime, most of
the previous features can be obtained in the weak coupling
limit  which has the advantage of often permitting
analytical computations where the physics is more transparent.

Superconductivity has been obtained in the purely repulsive model
with quantum Monte Carlo  \cite{QMC} and other numerical methods
\cite{otros}, with the antiferromagnetic fluctuation exchange
approximation \cite{flex1,flex2,flex3}, and with Renormalization Group
(RG) methods in the proximity of the Van Hove
singularity \cite{RG}. At present there is
no general agreement on the mechanism for
superconductivity. As for the symmetry of the superconducting order
parameter,   it is generally assumed to be d--wave in the hole
doped cuprates and remains unclear in the electron doped
compounds \cite{biswas}.

One of the most physically
appealing mechanisms for obtaining a pairing instability
from repulsive interactions goes back to Kohn and Luttinger (KL)
\cite{KL} who demonstrated the instability of a
three dimensional isotropic
Fermi system towards pairing due to the spatial
modulation of the effective interaction at a wavevector
of $2k_F$.  The KL mechanism has been
extended to other electron systems and dimensions
\cite{KC88,BCK92,C93} and has been extensively studied
in the 2D Hubbard model
\cite{BK92,CL92,ZS96,H99}. The Kohn-Luttinger mechanism
has very recently been reanalyzed in \cite{GS02}.

In the study of the instabilities of an electron system
a principal role is played by
the electron-hole susceptibility $\chi(\vec k , w)$ whose
imaginary part measures the density of electron--hole pairs
with energy $\omega$ and whose real part renormalizes the scattering
amplitude.  Scattering processes which involve opposite
points of the Fermi surface can be enhanced by its
special  geometry (nesting and Van Hove singularities are
extreme examples), or by other physical features.

In ref. \cite{us}, a scaling analysis was used to study the
pairing instabilities of a general Fermi surface in the 2D square
lattice as a function of its geometry. It was shown that the
curvature of the Fermi line modulates the effective interaction
in the BCS channel in such a way that
different harmonics scale as different powers of the scaling
parameter. As the latter goes to zero some harmonics become
negative giving rise to a KL superconductivity in the
given channel.

In this paper we will perform a KL analysis of electron
doped cuprates along the lines of ref. \cite{us} based
on special features of the susceptibility.
Recent experiments on electron doped cuprates\cite{biswas} propose a
change in the symmetry of the
superconducting order parameter from d wave below and around optimal
doping, to s wave in the overdoped regime.
Moreover, even when the gap has overall d-wave symmetry, its
angle-dependence evolves with doping, picking up
substantial harmonic content for hole underdoping, and there
is some evidence\cite{bgreen} that the peak shifts
away from $(\pi ,0)$ in electron-doped cuprates. This issue will also
studied in the paper.

The organization is as follows: In sect. II we set the model,
review the main arguments of the scaling analysis
of ref.\cite{us} and establish the d-wave nature of the superconducting phase.
In sect. III we analyze the
structure of the spin susceptibility to be used in the calculation.
Sect. IV is devoted to the evolution of the
symmetry of the order parameter with doping followed by our conclusions.

\section{The model and Kohn-Luttinger mechanism}

The Fermi surface of the electron compounds for the doping
values of interest has a general rounded shape
centered at $(\pi, \pi)$ with flatter portions in the diagonal
directions. The  the
t--t'--Hubbard model with
negative values of t' is the simplest model that reproduces
the observed feature. The dispersion relation is given by
\begin{equation}
\epsilon(\vec k)=-2t(\cos k_x+\cos k_y)-4t'\cos k_x\cos k_y \;.$$
\label{disprel}
\end{equation}
For definiteness we assume $t=0.326eV$, $t'/t=-0.276$ in our
calculations\cite{param,RMstr,PeAr,RoRi}.  

The Kohn-Luttinger mechanism is closely related to Friedel oscillations.
It is well known that, due to the sharp cutoff of the electron
distribution in $k$-space at the Fermi level, impurity potentials in a
metal do not fall off monotonically but have a superposed (Friedel)
oscillation.  Kohn and Luttinger showed that a similar oscillation arises
in the electron-electron interaction, leading to an {\it attractive}
interaction between two electrons separated by the right distance -- the
position of the first Friedel minimum.  In turn, the attractive
interaction can lead to a superconducting instability.  

The calculation can readily be formulated in renormalization group
language\cite{shankar}, based on the fact that the effective coupling
constants of a given hamiltonian (vertex functions) acquire an
energy-momentum dependence upon renormalization behaving like effective
potentials.  In the simplest Fermi liquid model (gas of electrons with
spherical Fermi surface and short range four fermi interactions),
only the forward and BCS channels get renormalized\cite{shankar}.
The standard KL mechanism occurs when the effective BCS vertex
at a given momentum ($2k_F$) oscillates in such a way that some of its
Fourier components becomes negative. The system then undergoes a
superconducting transition. The symmetry of the superconducting order
parameter can be found by expanding the potential  in eigenfunctions of
the symmetry of the model (spherical harmonics in the spherical case)
and finding the lowest negative eigenvalue.

This is the analysis that we will follow in sect IV of the paper
adapted to case the  Fermi surface given by the
contour lines of eq. (\ref{disprel}).

The KL mechanism for the rounded Fermi surface corresponding to electron doped
cuprates was analyzed in \cite{us} and shown to  induce a pairing
instability with $d_{x^2-y^2}$ symmetry
without special features of the susceptibility.
Based on very simple scaling arguments, it was shown that
the electron
susceptibility is proportional to $(1/f^2)$
where $f$ is the curvature of the Fermi surface
and gets modulated by it. Hence it has maxima for the scattering
vectors joining two opposite points of the
Fermi surface in the $(\pi/2, \pi/2)$ direction and equivalent points
where the curvature reaches its minimum
value. It has minima in the zero and equivalent directions.
This situation corresponds to $d_{x^2-y^2}$ symmetry. This analysis
is independent on the specific form of the spin susceptibility.
The detailed study of the susceptibility in the next section
reinforces the symmetry arguments of   \cite{us} and the
  $d$--wave character of the instability.

\section{The structure of the spin susceptibility}

 Recent angle-resolved photoemission spectroscopy (ARPES) of
 electron-doped Nd$_{2-x}$Ce$_x$CuO$_{4\pm\delta}$ (NCCO)\cite{nparm}
 found a smooth evolution of the band dispersion with doping, from half
 filling to optimal doping, which could be interpreted\cite{KLBM}
 in terms of the gradual filling of the upper
Hubbard band.  Doping gradually reduces the Mott gap that
closes near optimal doping.  This doping also falls
close to  another interesting point \cite{RM4} where the
displaced Fermi surface is tangent to the original one
and the susceptibility $\chi$ at $\vec Q =(\pi ,\pi )$
starts to drop precipitously.

Near this point, the magnetic susceptibility has the form of a nearly
flat-topped plateau in momentum space, with sharp falloff away from the
plateau.  The plateau is defined by the presence of points at which the
Fermi
surface (FS) overlaps the FS image shifted by $\vec Q$.  If the FS image is
shifted away from $\vec Q$ by an additional $\vec q{\ }'$, then for some
critical value $\vec q{\ }'=\vec q_c$, one or more of the  overlaps
ceases to exist as the two FS's pull apart.  The value of $q_c$, which
defines the plateau boundary in a given direction, satisfies
\begin{equation}
-2t(\sin (q_x/2)+ \sin (q_y/2))-4t'\sin (q_x/2)\sin (q_y/2)=\mu,
\label{eq:0e}
\end{equation}
where $\mu$ is the chemical potential.

\begin{figure}
\epsfxsize=0.33\textwidth\epsfbox{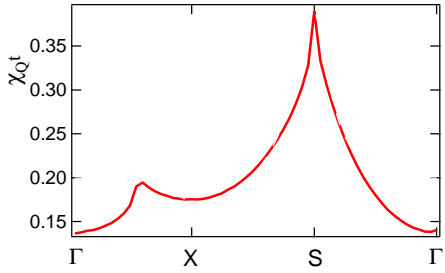}
\vskip0.5cm
\caption{Bare susceptibility at the termination of the plateau
described in the text, $\mu =0$. Brillouin zone points are $\Gamma$ =
(0,0), $X$ = $(\pi ,0)$, $S$ = $(\pi ,\pi )$.}
\label{fig:2}
\end{figure}

The plateau exists in the doping range $0\ge\mu\ge\mu_{VHS}=4t'$, where
$\mu_{VHS}$ is the doping of the Van Hove singularity.
When $\mu =0$, the width of the plateau shrinks to zero.
The plateau approximately satisfies the form $\chi =A-B\Theta (q'-q_c)
\sqrt{q'-q_c}$, with $\Theta$ a step function.  As the width
of the plateau vanishes $q_c\rightarrow 0$, the two square root terms
merge at a single point.  The resulting susceptibility
is depicted in Fig.\ref{fig:2}.  We find that the {\it angle
dependence} of the resulting superconducting gap function changes
dramatically depending on whether the doping lies on or off of this
susceptibility plateau (leading ultimately to changes in the gap
{\it symmetry}).

At T=0, the susceptibility $\chi_{\vec Q}$ can be written
\begin{eqnarray}
\chi_{\vec q}=\int{d^2\vec k\over (2\pi )^2}{1\over (\epsilon_{\vec
k+\vec q}-
\epsilon_{\vec k})}
\nonumber \\
={1\over 2t}\int{d^2\vec k\over (2\pi )^2}{1\over
(\cos {(k_x+q_x)}+\cos {(k_y+q_y)}-\cos k_x-\cos k_y)}.
\label{eq:C2}
\end{eqnarray}
The Fermi functions limit the integral to a sum of approximately
wedge-shaped areas.  We begin at the antiferromagnetic wave-vector, $\vec
q=\vec Q=(\pi ,\pi )$.
Letting $k_i=\pi /2+k'_i$, $i=x,y$, then to lowest order the energy
becomes
\begin{eqnarray}
\epsilon_{\vec k} =2\sqrt{2}tk'_{\perp}+2t'k_{\parallel}^{'2}
\label{eq:C3a}
\end{eqnarray}
with $k_{\parallel}$ and $k_{\perp}$ the momenta parallel and
perpendicular to the zone diagonal (magnetic Brillouin zone boundary).
Linearizing the energy denominator, $\Delta\epsilon\propto k_{\perp}$,
independent of $k_{\parallel}$
\begin{equation}
\chi_{\vec Q}\approx
{1\over 8\pi^2\sqrt{2}t}\int_0^{k_c}{dk_{\perp}dk_{\parallel}\over
k_{\perp}}={I\over 4\pi^2t}.
\label{eq:1}
\end{equation}

\begin{figure}
\epsfxsize=0.33\textwidth\epsfbox{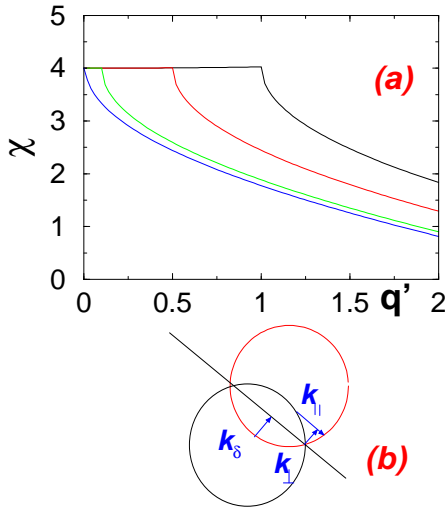}
\vskip0.5cm
\caption{(a) Calculated susceptibility $\chi (q)$ for several values of
overlap $k_{\delta}$ (from right to left, $k_{\delta}/q$ = 0.5, 0.25,
0.05, and 0, the last corresponding to the termination of hot spot
overlap). (b) Schematic of Fermi surfaces, defining $k_\delta$,
$k_{\perp}$ and $k_{\parallel}$.}
\label{fig:11b}
\end{figure}
The FS centered at $(\pi ,\pi )$ and the Q-shifted FS are illustrated
schematically in Fig.~\ref{fig:11b}b.
The region of integration is over the part of the upper FS in
Fig.~\ref{fig:11b}b not overlapped by the lower (Q-shifted) FS, and
$k_{\perp}$ ranges from zero at the apex of the wedge to $k_c=k_F-k_\delta$
at the middle of the upper FS, where $k_F$ is the radius of the FS and
$k_\delta$ is the overlap
parameter defined in Fig.~\ref{fig:11b}b.
Assuming $k_\delta <<k_F$ and keeping only the lowest order contributions,

\begin{eqnarray}
I=\sqrt{k_F}[\int_0^{k_c}dk_\perp
\frac{\sqrt{k_\perp + k_\delta}}{k_\perp}
-\int_0^{k_\delta}dk_\perp
\frac{ \sqrt{k_\delta-k_\perp}}{k_\perp}]  \nonumber \\
=
2k_F+\sqrt{k_Fk_\delta}\ln|\frac{1-\beta}{1+\beta}
|
,
\label{eq:C6b}
\end{eqnarray}
with $\beta =\sqrt{k_\delta/k_F}$.
For the $q$-dependent susceptibility, let $\vec q=\vec Q+\vec
q{\ }'$.  Then the FS is shifted by $\vec q{\ }'$, or $k_\delta\rightarrow
k_\delta+q'/2$ for the surface shown, $k_\delta\rightarrow k_\delta-q'/2$
for the
FS at $(-\pi ,-\pi )$, so $\chi_q\propto
I_{k_\delta+q'/2}+I_{k_\delta-q'/2}$.
There is one correction.  For $q'>2k_\delta$, the two FSs no longer
overlap, and the integral runs over the full half-circle, but $k_\perp$ is
measured from half way between the two FSs, so
$I_{k_\delta-q'/2}=2k_F[1-\gamma\tan^{-1}1/\gamma ]
$, with $\gamma =\sqrt{(q'-2k_\delta )/2k_F}$.  Figure~\ref{fig:11b}a
shows the
resulting susceptibilities for several values of $k_\delta$.  The
calculation explains the very flat top with weak positive curvature and
the sharp falloff at $q'=2k_\delta$.  For increased electron doping, the
hole-like Fermi surface shrinks, $k_{\delta}\rightarrow 0$, and the
plateau width shrinks to zero.  Near the plateau edge $\chi\sim 1-\pi\gamma
/2$ varies as $\sqrt{q'}$.  The cusp-like susceptibility in Fig. 2a at the
point where the plateau terminates corresponds to the $\chi_0$ peak in
Fig. 1 at $S=(\pi ,\pi)$. With the assumed parameter values, this
happens when $\mu =0$, corresponding to an electron doping $x\simeq -0.19$;
on the hole doped side, the plateau terminates when $\mu =4t'=-0.359eV$,
at a hole doping, $x=0.24$.

\section{Coupling Constant Calculations}

Calculation of the {\it magnitude} of the superconducting transition gap
is beyond the scope of the present paper: this requires a better
understanding of (a) the proper choice of susceptibility, (b)
incorporation of the frequency dependence of the susceptibility, (c)
proper accounting of the competition with magnetic ordering, and (d)
solution of the resulting (generalized) Eliashberg equations.

The most negative coupling
constant determines the dominant gap symmetry, while the
corresponding eigenfunction gives the angle dependence
of the gap.

In this section we investigate the
possible change in the symmetry of the superconducting order
parameter and how its angle-dependence
evolves with doping\cite{bgreen}.

The pairing coupling constant in a
given symmetry sector is given by the matrix \cite{flex2,H99}
\begin{equation}
\lambda_{n,m}^\alpha = \frac{1}{(2\pi)^2}\int\frac{d {\bf
k}}{v_k}\int\frac{d {\bf k'}}{v_{k'}}
V({\bf k}, {\bf k'}) \Delta_{\alpha n}({\bf k})\Delta_{\alpha m}({\bf k'})
\;\;.
\label{eq:6}
\end{equation}
where $V({\bf k}, {\bf k'}) =U+U^2\chi({\bf k}+{\bf k'})$, and
$\Delta_{\alpha n} ({\bf k})$ is the normalized ($\int \frac{d {\bf
k}}{v_k} \Delta_{\alpha m}\Delta_{\alpha n}({\bf k})=\delta_{mn}$) weight
function expanded in terms of the irreducible representations
$h_{\alpha n}$ of the symmetry group.  We approximate the CuO$_2$ plane
by a square lattice, in which case the appropriate symmetry group is
$D_4$, for which there are four singlet and one doublet representations.
These representations define gap symmetry sectors: while Eq.~\ref{eq:6}
can mix basis functions {\it within} a given sector, it does not mix
functions {\it between} sectors.  The four singlet sectors are labelled
according to their lowest basis functions, as $s$, $d$ (for
$d_{x^2-y^2}$), $d_{xy}$, and $g$, while the doublet sector is labelled
$p$ -- actually, there are two subsectors, $p_x$ and $p_y$ which do not
mix, but have degenerate eigenvalues.  The corresponding basis functions
are (with $n$ ranging from 0 to $\infty$)
\begin{eqnarray}
h_{s,n}(\phi) &=&\cos[4n\phi] \nonumber\\
h_{d,n}(\phi) &=&\cos[(4n+2)\phi] \nonumber\\
h_{d_{xy},n}(\phi) &=&\sin[(4n+2)\phi] \nonumber\\
h_{g,n}(\phi) &=&\sin[4(n+1)\phi] \nonumber\\
h_{p_x,n}(\phi) &=&\sin[(2n+1)\phi] \nonumber\\
h_{p_y,n}(\phi) &=&\cos[(2n+1)\phi].
\label{eq:7}
\end{eqnarray}
The above functions describe the orbital symmetry -- the singlet (doublet)
representations corresponding to spin singlets (triplets).  Below, we
evaluate the {\it lowest} eigenvalue for each symmetry sector, and label
the corresponding eigenfunction by the sector -- even though, e.g., the
lowest s-wave solution is really an `extended-s' solution, with all the
$h_{s,n}$, $n>0$, orthogonal to the pure s-wave solution.

For $\mu\le\mu_{VHS}$, the Fermi surface is an electron-like Fermi surface
closed about the $\Gamma$ point $(0,0)$, and the angle $\phi$ must be
measured about this point.  For $\mu\ge\mu_{VHS}$, the topology of the
Fermi surface changes to hole-like, centered on $(\pi ,\pi )$, and $\phi$
must now be measured from this point.

\begin{figure}
\epsfxsize=0.33\textwidth\epsfbox{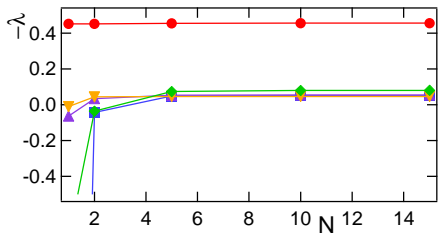}
\vskip0.8cm
\caption{Development of coupling constant with increasing matrix size,
$N$, for $\mu=-0.1eV$.  Circles = $d_{x^2-y^2}$; up-pointing triangles =
$g$; diamonds = $p_x$; squares = $s$; down-pointing triangles = $d_{xy}$.}
\label{fig:3}
\end{figure}
The maximal superconducting coupling is given by the minimal (i.e.,
maximally negative) eigenvalue of the $\lambda$ matrix, Eq.~\ref{eq:6}.
To solve this equation, the matrix was cut off at a finite size $N\times
N$, with $N= 15$.  This large $N$ value was employed to assure adequate
convergence in all sectors.  This is illustrated in Fig.~\ref{fig:3}.  If
the $N$=1 eigenvalue is already negative, the only change with $N$ is a
small increase in magnitude (due to level repulsion).  But if the $N$=1
eigenvalue is positive, $N$ plays a larger role.  There are two effects:
first, a diagonal matrix element might itself be negative and second, level
repulsion always pushes the largest and smallest eigenvalues away from the
mean.  For example, for the s-wave sector, all the diagonal elements are
found to be positive, but level repulsion generally leads to a small
negative eigenvalue.  However, for the parameter range studied, this is
always too weak to be of interest.

\begin{figure}
\epsfxsize=0.33\textwidth\epsfbox{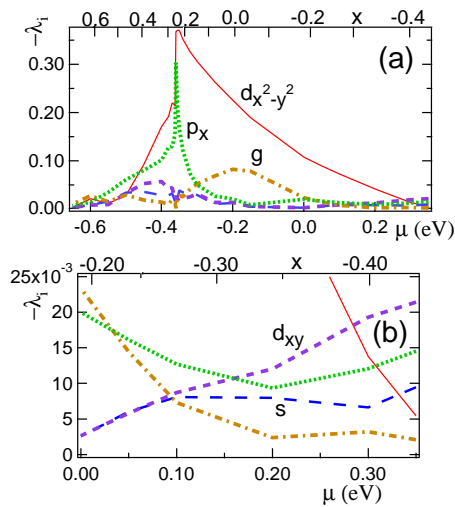}
\vskip0.8cm
\caption{(a) Evolution of the superconducting effective coupling as a
function of doping, for a variety of gap symmetries: solid line =
$d_{x^2-y^2}$; dot-dashed line = $g$; dotted line = $p_x$; long dashed =
$s$; short dashed = $d_{xy}$. (b) Blow-up of electron doping.}
\label{eigen}
\end{figure}
The results of the analysis are shown in fig. \ref{eigen}
for a variety of dopings (for electron doping $x$ is considered
negative), assuming $U=6t$.  Reasonable $\lambda$ values
are found for both electron and hole doping, but do not include the
suppression
of superconductivity near half filling caused by the magnetic order.  While
$\lambda$ decreases with electron doping, the preferred symmetry remains
d$_{x^2-y^2}$ over the doping range of interest.  The present
calculation thus provides no indication for a $d$ to $s$ crossover of the
symmetry.  However, at a higher doping, $x\sim -0.39$, there is a crossover
from d$_{x^2-y^2}$ to d$_{xy}$ symmetry; near such a crossover, there is
likely
to be a range of (gapless) d$_{x^2-y^2}$+id$_{xy}$ symmetry, which may
simulate an s-wave gap.  Alternatively, the symmetry change may be associated
with an additional pairing contribution due to electron-phonon coupling.
It is interesting to see that d-wave symmetry is dominant at both sides of
half filling, as found experimentally.

\begin{figure}
\epsfxsize=0.33\textwidth\epsfbox{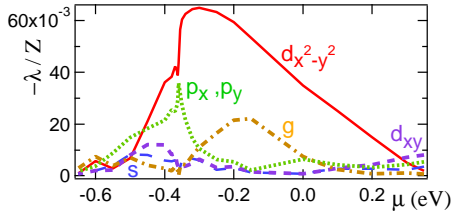}
\vskip0.5cm
\caption{Normalized coupling constants, $\lambda /Z$, for same data
as in Fig.~\protect\ref{eigen}.}
\label{fig:5}
\end{figure}

The Van Hove instability is clearly noticeable as the
(divergent) peak in the hole-like sector.  However, for superconductivity,
the relevant parameter is likely to be $\lambda /Z$, Fig.~\ref{fig:5},
where $Z=1+\lambda_s$ and $\lambda_s$ is the (0,0) matrix element
in the s-wave sector of Eq.~\ref{eq:6} (note
that this is the only term which includes the linear-in-$U$ contribution
to $V$).  This renormalization eliminates the Van Hove peak, shifting
the largest d-wave gap to a lower hole doping.  This is quite
suggestive of the experimental situation, where there appears to be a
quantum critical point somewhat above optimal hole doping\cite{Tall},
which may be associated with the Van Hove singularity (VHS)\cite{RM4}.

\begin{figure}
\epsfxsize=0.33\textwidth\epsfbox{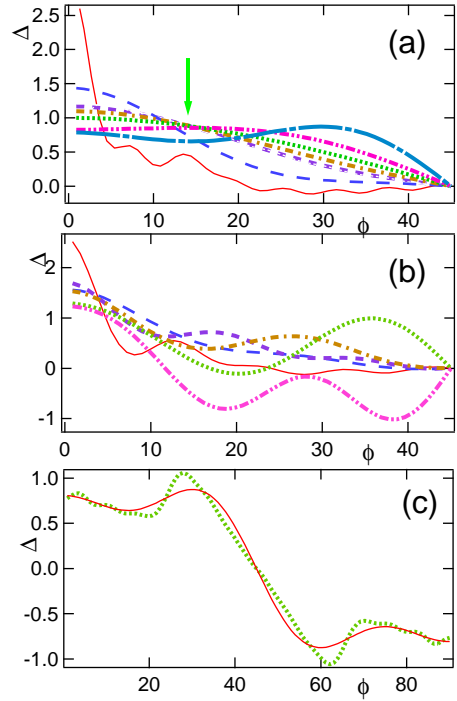}
\vskip0.5cm
\caption{(a) Angular dependence of the superconducting gap for the d-wave
symmetry solution, for several dopings: the chemical potentials [hole
densities] are $\mu$ [$x$] = -0.3599 [0.247] (thin solid line), -0.35
[0.22] (long dashed line), 0  [-0.19] (short dash-dotted line), 0.10
[-0.28] (dotted line), 0.20 [-0.35] (dash-dot-dotted line), and 0.3eV
[-0.41] (long dash-dotted line).
Arrow = datum of Ref.\protect\onlinecite{bgreen}.
(b) Continuation to higher hole doping, with $\mu$ [$x$] = -0.3599 [0.247]
(thin solid line), -0.38 [0.30] (long dashed line), -0.4 [0.33]
(short dashed line), -0.45 [0.41] (dot-dashed line), -0.5  [0.49]
(dotted line), and -0.55 [0.55] (dash-dot-dotted line).
(c) gap function for $\mu =0.3eV$, comparing a calculation employing all
15 harmonics (dotted line) with one involving only the dominant 4
harmonics (solid line).}
\label{fig:11}
\end{figure}

We find a striking evolution of the {\it angular dependence} of the
d--wave gap, as shown in Fig.~\ref{fig:11}.  The Kohn-Luttinger mechanism
leads to significant harmonic admixture, which changes as a function of
doping.  Including all harmonics, the gap functions contain excessive
structure, so the following approximation was introduced, to provide
smoother and more robust gap functions.  For each added harmonic order
(going from an $N\times N$ to an $(N+1)\times (N+1)$ matrix), the change
in the smallest (largest negative) eigenvalue was monitored, and if the
fractional change was less than some small reference value $\alpha$
(typically, $\alpha =0.02$ was used), the coupling to this harmonic was
neglected.  This reduced the matrix problem from $15\times 15$ to $N\times
N$, where $N$ was generally 2-4, except in the immediate vicinity of the
Van Hove singularity, where more harmonics were needed.  An example is
shown in Fig.~\ref{fig:11}c, while Figs.~\ref{fig:11}a,b show only
smoothed data, except at the VHS, where all harmonics are shown in
Fig.~\ref{fig:11}a, and only the dominant $N=7$ in Fig.~\ref{fig:11}b.

For all hole dopings the peak stays close to ($\pi ,0)$, but harmonic
content tends to sharpen the peak and flatten $\Delta$ near the nodes,
deviating from the simplest $\cos k_x-\cos k_y$ form.  Such flattening has
been found in both ARPES\cite{AD,Gol} and STM\cite{KMc} experiments.
However, the experimental trend is that the harmonic content is enhanced
in underdoped samples\cite{Din}, whereas the calculated trend is for larger
harmonic content to develop near the VHS -- i.e., with increasing hole
doping.  [In comparing to experiment, it must be kept in mind that, for
$\mu >-0.3599eV$, all angle $\phi$ measurements are centered at $(\pi
,\pi)$, with $\phi =0$ corresponding to $(\pi ,0)$.]

\begin{figure}
\epsfxsize=0.33\textwidth\epsfbox{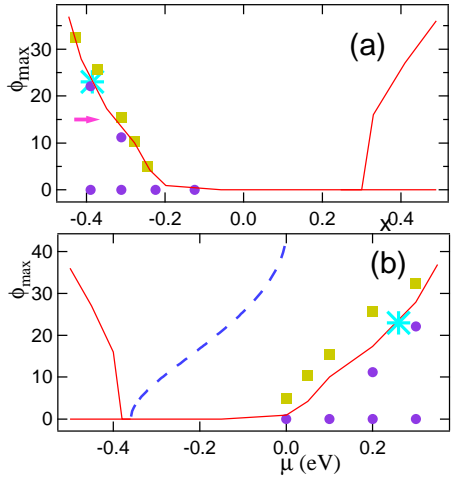}
\vskip0.5cm
\caption{Shift of $d_{x^2-y^2}$ peak from $(\pi ,0)$ as a function of
doping (a) and chemical potential (b).  The filled circles (squares)
correspond to changing $t'$ to $-0.12t$ ($-0.425t$); the dashed line in
(b) is the position of the hot spots. Arrow in (a) = datum of
Ref.\protect\onlinecite{bgreen}.  Asterisk = point of crossover to
$d_{xy}$-symmetry.}
\label{fig:12}
\end{figure}

As the doping shifts toward electron-like, there is a significant shift of
the peak position with $\phi$, away from $(\pi ,0)$.  This result is
consistent with recent observations by Blumberg, et al.\cite{bgreen}
(One should, however, note the debate\cite{VHM,BG2}.)
However, the agreement is not quantitative: a shift of the peak to $\sim
15^o$ is found for an experimental doping of $x=-0.15$, whereas the
predicted doping is $-0.32$, Fig.~\ref{fig:12}.  Indeed, there appears to
be a close correlation between the shift of the d-wave peak and the
crossover of the symmetry from $d_{x^2-y^2}$ to $d_{xy}$: the crossover
occurs when the peak has shifted about half way to 45$^o$ (asterisk In
Fig.~\ref{fig:12}).  Note that a similar shift arises on the
hole-overdoped side, although a second, larger peak remains at $\phi =0$.

This shift of the peak from $(\pi ,0)$ does not follow the position
of the hot spots on the Fermi surface (dashed line in Fig.~\ref{fig:12}b),
but depends on whether or not the system is on the susceptibility plateau.
The peak stays close to $(\pi ,0)$ as long as the chemical
potential is on the hot spot plateau ($\mu <0$), then rapidly shifts
toward $45^o$.  Similarly, for $\mu <-0.3599eV$ (off of the other side of
the plateau, beyond the VHS), a second peak appears and shifts to $45^o$
by $\sim\mu =-0.5$, Fig.~\ref{fig:11}b, by which point the gap has crossed
over to $p$-wave symmetry.

For the electron doping case we have studied how this peak shift changes when
the plateau width (or $\tau$) is varied.  From Eq.~\ref{eq:6} the angle
dependence is controlled by the product of two terms, a weighted dos,
$g(\bf k)=\Delta_{s,1}(\bf k)k^2(\phi )/v_k$, and a weighted
susceptibility
$\bar V(\bf k)=\int d\phi'V(\bf k,\bf k')g(\bf k')$.  The latter $\bar
V(\bf k)$ peaks at $(\pi ,0)$ on the plateau, and starts shifting toward
$45^o$ as soon as $\mu$ is off of the plateau.  However, $g(\bf k)$
continues to peak at $(\pi ,0)$, and the product shifts off of $(\pi ,0)$
more slowly as $\tau$ is reduced.

In conclusion by rather general symmetry arguments
we have shown that d-wave superconductivity is a robust feature
of the cuprates both hole and electron doped. We have also examined
the evolution of the shape of the order
parameter with doping and found
a deviation of the order parameter angle dependence
from simple $\cos{2\phi}$-form similar to
that measured in recent experiments (Fig.
4).
Although the calculations are based in a weak coupling analysis we
believe that they are justified in the electron--doped case
and that more refined computations will not change the
general features.

Note added: Recently, we became aware of a similar calculation\cite{KYZ}.
Here, a doping-independent nearly-antiferromagnetic Fermi liquid
susceptibility was introduced in place of the lowest-order (Kohn-Luttinger)
form we assumed.  A very similar crossover of gap symmetry with electron
doping was found, with two differences.  First, the crossover was found
to be at the doping at which the hot-spot plateau terminated.
This is probably not very significant, however; the parameters are so
different that this actually corresponds to a higher doping ($x\sim -0.59$)
than we found, and much higher than in the experiments.  More significant
is that we find a crossover from $d_{x^2-y^2}$ to $d_{xy}$
symmetry, while their crossover is to $p$-wave symmetry.  We have repeated
our calculations using the full self-consistent susceptibility\cite{RM4},
and find for this nearly divergent susceptibility that the crossover is to
a state of either $p$- or $g$-wave symmetry.  These results will be
reported in a future publication.

{\it Acknowledgements}.
The financial support of the CICyT (Spain), through
grants no. PB96-0875,
1FD97-1358 and BFM2000-1107
is gratefully acknowledged. One of us (RM) has been supported
by the Spanish Ministerio de Educaci\'on through grant SAB2000-0034.


\begin{references}
\bibitem{param}These parameters have been found to fit the electronic
dispersion in undoped\cite{RMstr} and electron-doped\cite{KLBM} cuprates,
while very similar parameters fit the spin wave dispersion\cite{PeAr,RoRi}
in La$_2$CuO$_4$.
\end{references}
\end{document}